\documentclass[a4paper,11pt]{article}
\usepackage[dvipsnames]{xcolor}
\usepackage{pos}

\title{Dark Matter at ICRC 2023}
 \ShortTitle{Dark Matter rapporteur}

\author*[a]{Madeleine J. Zurowski}

\affiliation[a]{Department of Physics,\\
  University of Toronto, Toronto, ON M5S 1A7, Canada}

\emailAdd{madeleine.zurowski@utoronto.ca}

\abstract{This proceedings summarises the dark matter presentations at ICRC 2023. It aims to not only act as a reference document reporting the various results and projections, but also compares the different search methods and attempts to assess the complementarity of the experimental methods discussed.}

\ConferenceLogo{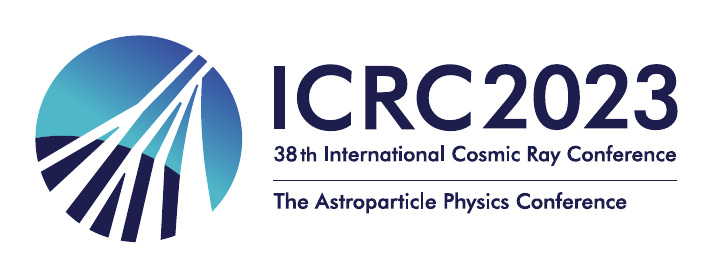}

\FullConference{%
38th International Cosmic Ray Conference (ICRC2023)\\
  26 July - 3 August, 2023\\
  Nagoya, Japan}

\begin{document}
\maketitle

\section{Introduction and content breakdown}
\begin{figure}[!h]
    \centering
    \includegraphics[width=0.8\textwidth]{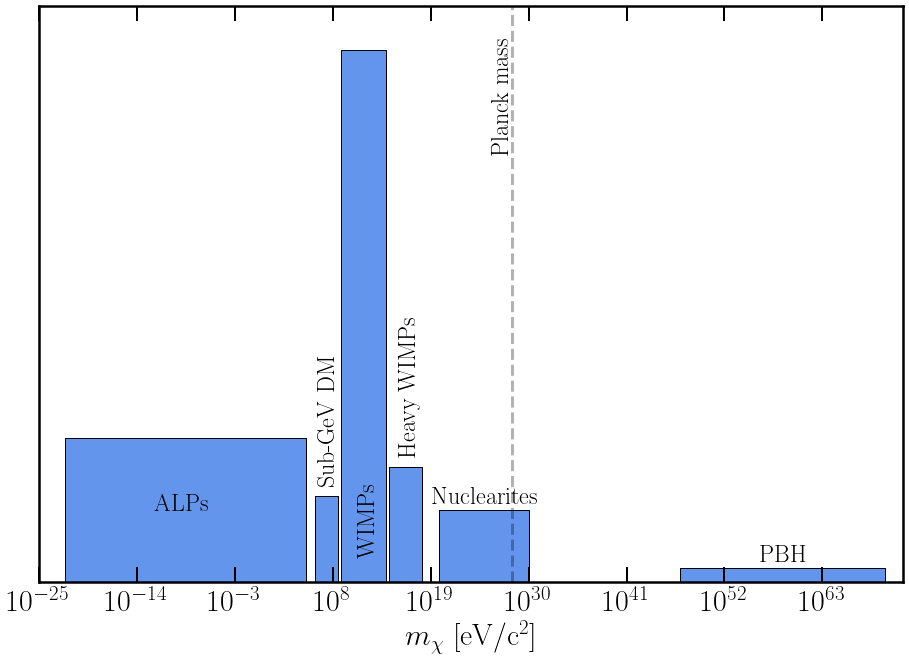}
    \caption{DM models discussed at ICRC 2023 as a function of mass.}
    \label{fig:model-break}
\end{figure}

Dark matter (DM) is a well motivated and widely studied problem in particle physics. Although there is an array of astrophysical evidence for DM, it has yet to be explicitly detected. Various DM models and methods of studying them are presented in Ref. \cite{calore}. The models discussed at ICRC 2023, weighted by the number of contributions they were mentioned in, are shown in Fig. \ref{fig:model-break} as a function of candidate mass. As might be expected, most contributions focused on WIMP or ALP DM; although there are some candidates above the Planck mass, in general the resulting rates are highly suppressed and so difficult to constrain experimentally.

Of the $\sim$80 contributions at the conference, 51.2\% focused on indirect detection, 32.6\% on direct detection, and the other 16.3\% on theory or other detection methods. This proceedings will summarise these contributions, beginning by comparing the observables for indirect and direct searches.

Indirect detection experiments probe annihilation and/or decay of DM. Both of these processes can produce a flux of observed particles $\frac{d\Phi_s}{dE}$, where $s$ is typically neutrinos or gammas. These fluxes are given by
\begin{equation}
\begin{split}
   \frac{d\Phi_{{\rm ann},s}}{dE} &= \frac{1}{4\pi}\frac{\langle\textcolor{blue}{\sigma_{{\rm ann}}} v \rangle}{2\textcolor{blue}{m_{\chi}}^2}\textcolor{teal}{\frac{dN_s}{dE}}\textcolor{LimeGreen}{\int_{\rm los}\rho^2(\vec{r})ds},\\
   \frac{d\Phi_{{\rm dec},s}}{dE} &= \frac{1}{4\pi}\frac{1}{2\textcolor{blue}{m_{\chi}\tau}}\textcolor{teal}{\frac{dN_s}{dE}}\textcolor{LimeGreen}{\int_{\rm los}\rho(\vec{r})ds}.\\
\end{split} 
\end{equation}
Direct detection, on the other hand, probes the scattering of DM with protons, neutrons, or electrons, which will produce an excess rate in the energy spectrum of a detector according to
\begin{equation}
    \frac{dR}{dE}=N_T\frac{\textcolor{LimeGreen}{\rho}}{\textcolor{blue}{m_{\chi}}}\textcolor{LimeGreen}{\int_{v_{\rm min}}^{v_{\rm esc}}vf(v)}\textcolor{teal}{\frac{d\sigma}{dE}}\textcolor{LimeGreen}{d^3v}.
\end{equation}
Both search methods depend on \textcolor{teal}{particle physics interactions} and \textcolor{LimeGreen}{astrophysical parameters} (as well as detector performance such as efficiency or resolution, which are not accounted for in these equations). As such, assumptions must be made about these terms in order to place constraints on the terms in \textcolor{blue}{blue} that tell us about the nature of DM; its annihilation cross section $\sigma_{\rm ann}$, decay lifetime $\tau$, mass $m_{\chi}$, and scattering cross section $\sigma$. Possible improvements or more cautious model building approaches that might impact these expressions will be laid out in Sec. \ref{assump}, before discussing the various search methods explored at this conference.

\section{Assessing assumptions}
\label{assump}
Clearly, there are a large number of theoretical modelling choices and assumptions that play into the calculations used to constrain DM properties at detectors. Not only will a clear, careful understanding of these improve any limits set, but they are also important when comparing results from different DM detection methods. For example, typically indirect detection experiments need to assume some decay scheme for the mediator particles in order to calculate the expected observation rate. However, the limits from one scenario (e.g., $\chi\chi\rightarrow\tau^+\tau^-$) will not necessarily constrain another (e.g., $\chi\chi\rightarrow b\bar{b}$), as there is no guarantee DM will couple to both. Direct detection however, makes no assumption on the mediators decay, and so it is not trivial to compare the constraints from the two methods. Put another way, $\sigma_{\rm ann}\neq\sigma_{\rm SI}$. Some of these assumptions, their validity, and possible improvements, were discussed at ICRC and are outlined here.\\

{\it Velocity distribution}\\
For direct detection searches, the velocity distribution $f(v)$ contains information about where the DM is in the galaxy, and how fast it is travelling. Typically, the assumption is made that it follows a Maxwell Boltzmann distribution (called the Standard Halo Model), though studies in the last few years have demonstrated that there may be deviations to this due to galactic substructure. Thus, it may be preferable to at times avoid making any assumptions and instead adopt a halo independent approach. In general (as might be expected), this relaxes the exclusion plots for direct detection \cite{skang}.\\

{\it J-factor}\\
Like velocity distribution, the J-factor, $\int_{\rm los}\rho^2(\vec{r})ds$, describes how DM is distributed along the line of sight between a telescope or detector and the target it is focused at. Three different J-factors were examined across various indirect detection searches - Ando, Geringer-Sameth, and Bonnivard. The impact of these varied depending on the target under examination, but in general the limits can change significantly (up to a factor of 10), where the Ando J-factor (which predicts the lowest amount of DM in the line of sight for most targets) produces the `worst' limits and the Bonnivard J-factor (which predicts more DM in the line of sight) are the strongest \cite{mcgrath,kersz}. This demonstrates the systematic uncertainty surrounding this value, and the need to assess the assumptions made when comparing analysis from different collaborations.\\

{\it Galactic Centre Excess}\\
There is clear agreement within the field that there is an excess of GeV gamma rays above known astrophysical backgrounds in the Galactic Centre. However, there is not currently clear agreement as to what could be causing this, and there are a number of competing views as to an explanation. It may be possible that more properly modelling the background uncertainty by using machine learning could make the situation clearer, though there is the caveat that there is ``only one sky'', and so the fact that reality is not part of the background model training data may make the results from any ML model untrustworthy \cite{eckner}.\\

{\it Modelling electromagnetic cascades}\\
A large number of ALP searches are done by assuming some conversion between ALPs and photons. As DM travels large distances across the cosmos to reach Earth, it travels through magnetised environments, producing possible measurable signals in the form of electromagnetic cascades. Properly modelling these effects can increase the expected energy spectrum and so may impact future searches \cite{batista}.\\

{\it Subhalo survival}\\
Galactic subhalos make excellent candidates for DM searches due to their large annihilation flux. However, it is still unclear whether we would expect these subhalos to survive to present day. Understanding this is a necessity if this phenomena is to be used in DM searches. Based on numerical simulations, the DM density profile gets truncated as the mass loss takes place. However, based on this study most subhalos will last until present day, even where most (99\%) of the mass is lost at accretion \cite{aguirre}.\\

{\it Annihilation cross section}\\
The total annihilation cross section, computed by accounting for the branching fraction of DM decay to all possible intermediate states, is not well constrained, largely due to the neutrino channel. Put another way - there is no way to compute the expected flux of DM secondaries in a model independent way. Better constraints on the neutrino cross section, coming with future neutrino detectors at KM3Net, can help to close (or uncover new physics in) the WIMP window \cite{ngnu}.\\

{\it Non-standard models}\\
Finally, perhaps there are non-standard models that have been overlooked with these typical assumptions for DM searches. For example, a modified gravity model can be shown to explain the Hubble Tension as well as some DM observations \cite{desimone}. There are also still scalar DM models that are valid and could be probed by the next generation high luminosity LHC \cite{avila}.

\section{Direct detection}
\begin{figure}[!h]
    \centering
    \includegraphics[width=0.95\textwidth]{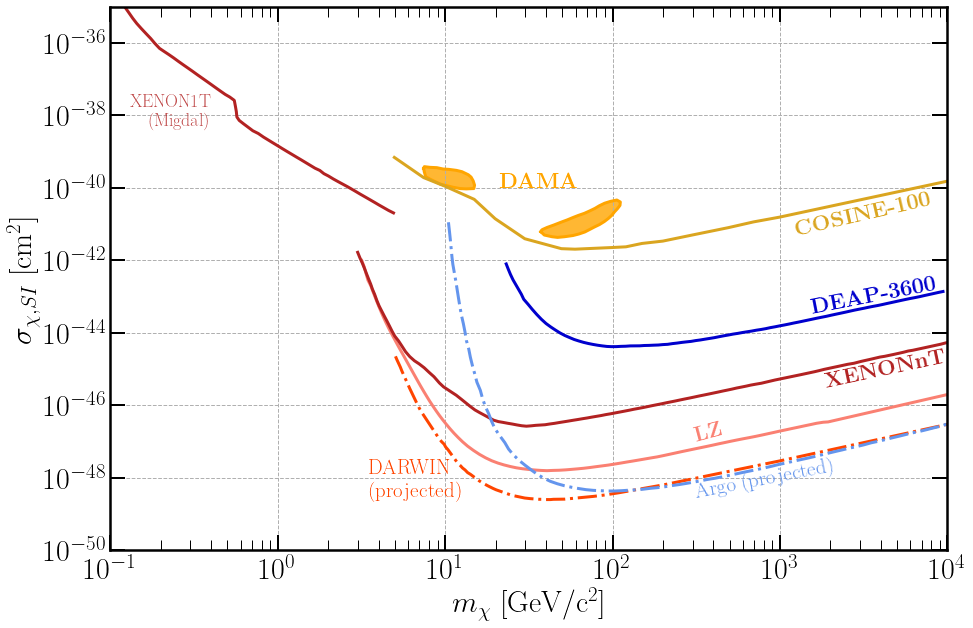}
    \includegraphics[width=0.95\textwidth]{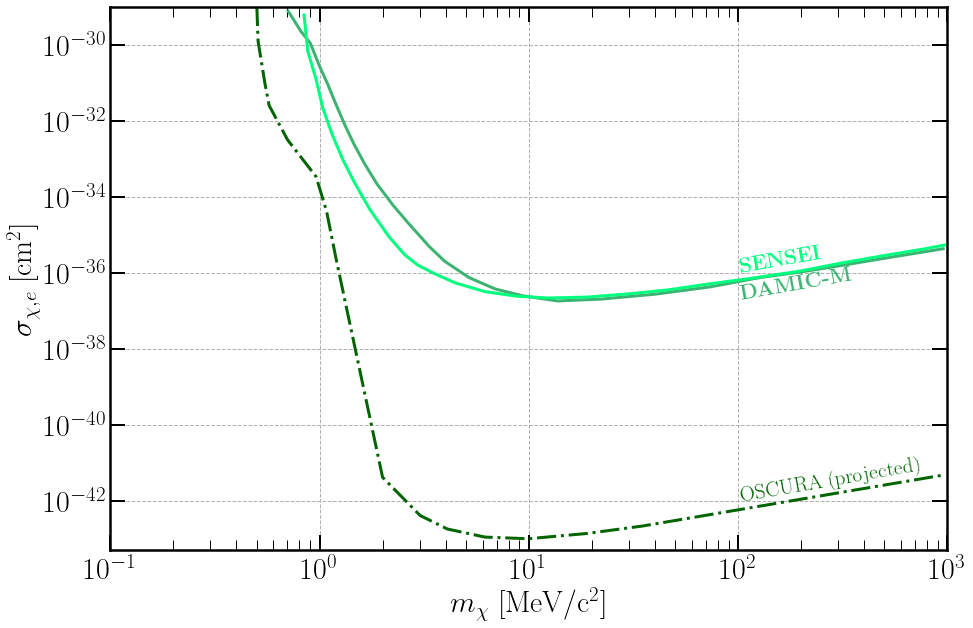}
    \caption{Constraints on spin independent nucleon scattering with equal proton and neutron coupling (top) and electron scattering with form factor 1 (bottom). Dashed lines indicate projected limits for future experiments. For the exclusion limits in the top figure, reddish lines indicate a Xe based experiment, blueish an Ar based experiment, and yellowish a NaI(Tl) based experiment. COSINE limits are from Ref. \cite{yu}, DEAP-3600 from Ref. \cite{tardif}, XENONnT from Ref. \cite{brown}, LZ and DARWIN from Ref. \cite{wang}, ARGO from Ref. \cite{walczak}, SENSEI and OSCURA from Ref. \cite{botti} and DAMIC-M from Ref. \cite{gaior}.}
    \label{fig:dd-lim}
\end{figure}

Direct detection searches probe the scattering of DM off targets through one (or some combination of) the following signal types: scintillation, ionisation, and phonons. Detectors are categorised based on this signal, where different methods will have different strengths and weaknesses. Those presented at ICRC are summarised in Table \ref{tab:dd}, with limits shown in Fig. \ref{fig:dd-lim}. They are also discussed in the following subsections, though it should be noted this is not a conclusive summary of full complement of the existing direct detection experiments or strategies.\\

\begin{table}[!h]
\centering
\begin{tabular}{l|l|l|l|l}
Experiment & Detector type & Material & Detection channel & Status \\ \hline
ARGO & TPC & Ar & Light and charge & Planned \\
COSINE & Crystal scint. & NaI(Tl) & Light & Data taken 2016-2023 \\
DAMIC-M & CCD & Si & Charge &  Commissioning\\
DarkSide & TPC & Ar & Light and charge & Upgrade under commission \\
DARWIN & TPC & Xe & Light and charge & Planned \\
DEAP-3600 & Liquid scint. & Ar & Light & Operating since 2016\\
LZ & TPC & Xe & Light and charge & Operating since 2022 \\
NEWAGE  & Directional TPC & CF$_4$ & Light and charge & Planned \\
OSCURA & CCD & Si & Charge & Planned \\
PICOLON & Crystal scint. & NaI(Tl) & Light & Planned \\
SABRE & Crystal scint. & NaI(Tl) & Light & Commissioning \\
SENSEI & CCD & Si & Charge & Operating since 2019\\
XENONnT & TPC & Xe & Light and charge & Operating since 2021\\
\end{tabular}
\caption{Direct detection experiments discussed at ICRC.}
\label{tab:dd}
\end{table}

{\it Crystal scintillators}\\
Crystal scintillator detectors used for DM detection are typically made of NaI(Tl). Their goal is to understand the longstanding DAMA modulation, which viewed by itself is consistent with a DM presence within the galaxy, but in the context of all other direct detection experiments, has been ruled out as DM under the standard WIMP paradigms. However, because comparing results from different targets requires the assumption of some interaction model, it is impossible to conduct a truly model independent test without NaI(Tl) detectors. COSINE (operating since 2016), SABRE (commissioning), and PICOLON (planned) are designed to provide such a test.\\
COSINE has presented results from their first three years of data taking with 100 kg of NaI(Tl) in South Korea, reporting a modulation that is consistent with both the DAMA rate and the null hypothesis. Their full six year run has been completed, and the collaboration is in the process of upgrading to lower background crystals with a higher light yield to reduce their energy threshold \cite{yu}. As well as the traditional modulation search (where the background is modelled as a sum of decaying exponentials plus some constant rate), COSINE have presented an analysis that mimics that conducted by DAMA, where the background is assumed to be flat and so the average rate over a year is subtracted from the data. It is well established that such a method will induce a sawtooth signal that can appear as a modulation \cite{buttazzo,messina}, but these results demonstrate that for the expected time profile of a NaI(Tl) experiment (decreasing with time) this analysis will produce a signal with the opposite phase, and so is unable to explain DAMA without a detailed understanding of their particular background model.\\
SABRE is a NaI(Tl) experiment in the construction phase. Its key feature is that it boasts two underground detector sites: one in LNGS, Italy, and the other in SUPL, Australia. By placing a detector in each hemisphere SABRE will be able to decouple any seasonally modulating backgrounds from that of the DM signal. The collaboration has produced the lowest background large scale crystals ($>$ 3 kg) since DAMA and is due to start taking data in the coming year \cite{mews}.\\
PICOLON is another planned NaI(Tl) that will be located in Japan and start taking data in 2025. They have pioneered new purification strategies for NaI(Tl) crystals (recrystallisation and resin filtering) that reduce the background of the experiment to below 0.1 cpd/kg/keV for smaller scale crystals, the lowest background of any NaI(Tl) collaboration \cite{fushimi,kotera}.\\

{\it Noble gases/liquids}\\
Noble elements are typically used at scintillation only experiments, or operated in a dual phase time projection chamber (TPC) where both light and charge are collected. The double signal of the TPCs lends itself well to particle identification and thus separating background from signal. Due to their relatively high target mass and large scale experiments, noble element based experiments set the strongest limits in the typical WIMP phase space. Presentations at ICRC focused on the Argon based DEAP-3600 and DarkSide, and the Xenon based LZ and XENONnT.\\
DEAP-3600 is a scintillation only, liquid Ar experiment based at SNOLAB, Canada. By developing new pulse shape discrimination strategies, the collaboration has improved signal to background separation. As well as this they have probed a wide array of different DM models, examining sensitivity to ultra-heavy WIMPs, a variety of effective field theory interactions, and different additions to velocity substructure in the galaxy \cite{tardif}.\\
DarkSide is a 20 tonnes dual phase Ar TPC planned for LNGS. It is currently in the process of being upgraded to start taking data in 2027, and is projected to set limits 10 times stronger than the previous DarkSide iteration. The ultimate goal of the Ar community is a collaborative experimental effort called ARGO that would be constructed from 300 tonnes of ultra pure Ar \cite{walczak}.\\
LZ is a Xe based TPC that currently boasts the strongest limits in standard WIMP phase space. The detector started taking science data in 2022 with 7 tonnes of liquid Xe. As well as exploring WIMP phase space, the physics program includes plans for examining the Migdal effect, effective field theories, more exotic DM models, astrophysical neutrinos, and rare $\beta$ decays \cite{wang}.\\
XENONnT is an upgrade of XENON1T including a neutron veto and a Radon distillation strategy to reduce background, the latter of which was critical for understanding the excess reported by XENON1T \cite{eising,kobayashi}. A few WIMP-like events were observed in the first set of data taking, but this is still consistent with the background model. It is also worth noting that XENONnT using a more conservative statistical approach than LZ, which is part of the reason the latter's limits are significantly stronger \cite{brown}.\\
As well as the two main Xe based experiments, there was significant, impressive R\&D based work for future Xe detectors, including a facility testing some of the electronic apparatus requirements for new larger scale detectors \cite{gaior}, new low background photo-sensor technology \cite{hasegawa,sakamoto}, and the use of TPCs to help characterise low background setups \cite{ito}. This is largely geared towards supporting DARWIN and/or the global XLZD consortium \cite{wang}.\\

{\it Charge coupled devices}\\
Charge coupled devices (CCD's) are used in DM detection, in particular for DM-electron scattering. When DM hits the Si devices it excites the electrons into the conduction band, and the resulting electron hole pairs are collected at pixels using an applied electric field. Skipper CCDs are a development of this technology that involves multiple reading from the same pixel (as opposed to measuring the charge in each pixel only once). This significantly reduces the readout noise and thus the background in these detectors \cite{botti}. In particular, this detector type promises the lower energy threshold in direct detection, and as such requires R\&D efforts to ensure that nuclear recoils are modelled correctly at these low energies \cite{erhart}. The detectors making use of this technology that were discussed at ICRC are SENSEI and DAMIC-M. \\
SENSEI ran a small, gram scale detector from 2019-2020 in Fermilab, USA before beginning commissioning at SNOLAB where it is currently taking data working towards a 100 g detector. The previous location gave the collaboration the opportunity to constrain millicharged DM (as well as the typical scattering) using the NuMI beamline \cite{botti}.\\
DAMIC-M is the most recent iteration of the DAMIC collaboration working towards installing a kg-scale Si Skipper CCD at Modane in France. They have published their first science results using a 20 g detector with a background of 10 counts/day/kg/keV \cite{damic}.\\
As well as these two separate experiments, there are plans for a combined detector from the two collaborations, OSCURA, that increases the mass by an order of magnitude. This search would operate with a 10 kg payload in 2027 \cite{oscura}.\\

{\it Novel detection strategies}\\
As the present array of DM direct detection experiments have seen no DM signal, and are fast approaching the neutrino fog, a number of alternative detection strategies and new interaction models have been proposed.\\
In particular, directional detection has been suggested as a method to carve through the neutrino fog. As the DM should be reaching Earth from the direction of the Cygnus constellation, particle tracking would allow for these signals to be distinguished from neutrino-nucleus scattering. NEWAGE is a proposed low pressure gaseous TPC filled with CF$_4$ gas that would be placed at the Kamioka mine. A small scale iteration ($\sim$0.04 m$^3$) has already demonstrated the feasibility of this technology. A larger upgrade is currently being constructed in collaboration with the Cygnus group \cite{higashino}.\\
Nuclearites, or macros, are another DM candidate that could be observed directly. These have masses on the scale of 10$^{30}$ eV, and so could interact directly at neutrino observatories \cite{paun}. They may also be large enough to view in the sky as long, fast moving meteor events. Two projects have been designed to detect such events; DIMS \cite{shinozaki,mori,kajino} and Mini-EUSO \cite{casolino}.\\
Similar to the concept of using neutrino detectors to observe much larger mass DM, they could also be used to test for boosted DM. Low mass, but high speed DM (that has been `boosted' to these higher speeds) would produce low energy electron recoils that are peaked towards the galactic centre. Such a signal is detectable at experiments such as SuperKamiokande \cite{iovine}. Other lighter DM particles (MeV scale) like a sterile neutrino could be undetectable at typically direct detection experiments, but observable at neutrinoless double beta decay setups \cite{nozzoli}.

\newpage
\section{Indirect detection}
Indirect detection experiments tend to be dominated by telescopes pointed at celestial objects. Common targets for DM searches are the galactic centre (centre), galaxy clusters (clusters), the Sun, and dwarf spheroidal galaxies (dSph). A summary that compares the main telescopes presented at ICRC is given in Table \ref{tele-tab}, and in the following subsection. The results are discussed based on the target, to understand how the slightly different sources and telescopes might be complementary, rather than looking at the results from each telescope in isolation.

\begin{table}[!h]
\centering
\begin{tabular}{l|l|l|l|l}
Telescope & Energy [GeV] & Source & Targets & Status \\ \hline
ANTARES & 10 - 10$^5$ & Neutrino & Centre, Sun & Operated 2007-2022 \\
CTA & 10$^2$ - 10$^5$ & Gamma & Clusters, dSph & Planned \\
FAST & sub eV & Radio & dSph & Operating since 2020 \\
Fermi-LAT & 2$\times$10$^{-2}$ - 10$^3$ & Gamma & Centre, Sun, dSph & Operating since 2008 \\
HAWC & 300 - 10$^5$ & Gamma & Centre, dSph & Operating since 2015 \\
HESS & 30 - 10$^5$ & Gamma & Centre, dSph & Operating since 2003 \\
IceCube & 3$\times$10$^3$ - 6$\times$10$^5$ & Neutrino & Centre, clusters & Operating since 2010 \\
KM3Net/ARCA & 10$^2$ - 10$^8$ & Neutrino & Sun & Commissioning \\
KM3Net/ORCA & 1 - 10$^2$& Neutrino & Centre & Commissioning \\
LHAASO & 10$^2$ - 10$^6$ & Gamma & dSph  & Operating since 2019 \\
MACE & 20 - 10$^4$ & Gamma & dSph & Operating since 2020 \\
MAGIC & 50 - 5$\times$10$^4$ & Gamma & Centre, dSph & Operating since 2004 \\
SWGO &  10$^2$ - 10$^5$ & Gamma & Centre, dSph & Planned \\
VERITAS & 85 - 3$\times$10$^4$ & Gamma & dSph & Operating since 2007 \\
\end{tabular}
\caption{Telescopes used for indirect DM searches presented at ICRC. Note that the targets for each telescope are not intended to be a complete list, only those for which results were discussed at the conference.}
\label{tele-tab}
\end{table}

\subsection{Telescopes}
{\it ANTARES}\\
ANTARES was a neutrino telescope located in the Mediterranean Sea 25 km under water, operated from 2007 to February 2022, collecting 4532 days of data. It was constructed from 12 vertical lines, each containing 75 Optical Modules. Data cuts for analysis were chosen to optimise for the discrimination of DM over atmospheric neutrinos, and work is ongoing to develop machine learning algorithms for low-energy event reconstruction \cite{garcmend}. All data analysed to date is consistent with background only models \cite{gozzini}.\\

{\it CTA}\\
CTA is a planned telescope consortium, which will be made up of two Cherenkov telescope arrays in the Northen and Southern hemisphere. The former will focus on the low energy region, while the latter will look at galactic sources. The Northern array will be located in Spain and made up of 4 large telescopes and 9 medium sized ones, while the Southern array in Chile is made up 14 medium sized and 37 small sized telescopes.\\

{\it FAST}\\
FAST is a filled aperture, single dish radio antenna with a diameter of 500 m located in China. It was comissioned in 2016 and became operational in 2020. It has a field of view of 40$^{\circ}$ that does not include the Galactic Centre. Unlike the other telescopes discussed here, FAST searches for radio emissions rather than gammas or neutrinos, and so probes a very different energy region.\\

{\it Fermi-LAT}\\
Fermi-LAT is a satellite that has been in orbit at and altitude of 550 km since 2008. The detector module is an array of towers that contain trackers and calorimeters. It has a field of view of around 20\% of the sky, and is swept in such a way that it is able to scan the whole sky in approximately 3 hours (two orbits) \cite{kersz}.\\

{\it HAWC}\\
HAWC is an array of water Cherenkov detectors located in Mexico at an altitude of 4.1 km with a field of view of 15\% of the sky \cite{kersz}. It began operating with its full main array in 2015, but a recent hardware expansion is expected to improve sensitivity in the high region by a factor of 2-4. As well as probing the typical WIMP-scale DM, it is able to test for Lorentz invariance violation as well as primordial black holes and axion like particles \cite{harding}.\\

{\it HESS}\\
HESS began as an array of 4 Cherenkov telescopes with a diameter of 12 m each in 2003 (phase I), and was upgraded to Phase II in 2012 with the addition of a telescope that has a diameter of 28 m. It is located at an altitude of 1.8 km in Namibia and has a 5$^{\circ}$ field of view \cite{kersz}.\\

{\it IceCube}\\
IceCube is a neutrino detector located in the ice at the South Pole. It features 5,160 digital optical modules distributed on 86 ``strings'' that reach down to the depth of 24500 m. The experiment also includes a 1 km$^2$ surface array made up of 324 optical modules. The detector was completed and began operation in 2010.\\

{\it KM3Net}\\
KM3Net is made up of two neutrino telescope sites at the bottom of the Mediterranean Sea: ORCA (off the coast of France) and ARCA (off the coast of Italy). The two are designed to detect different energy scale events, with ORCA optimised for the lower energy region ($<$100 GeV) and ARCA the higher ($>$100 GeV). At present, ORCA consists of 18 detection units (collections of photomultiplier tubes), while ARCA is made up of 21. The analysis presented at ICRC uses only a subset of these (8 for ORCA and 6-8 for ARCA) for $\sim$500 days of live time, as it is based on data taken during commissioning \cite{saina, gutierrez}. It is worth noting that even with this small subset of the data that will ultimately exist, KM3Net is already able to place limits competitive with experiments with a much longer lifetime.\\

{\it LHAASO}\\
LHAASO is gamma ray telescope located in China. It is made up of 3 subdetector arrays giving it the ability to reach energies as high as 1 PeV. These arrays are the KM2A (kilometre square array), WCDA (water Cherenkov detector array) and the WFCTA (wide field of Cherenkov telescope array). The DM results presented at ICRC 2023 are largely from the KM2A data \cite{li}.\\

{\it MACE}\\
MACE is a Cherenkov telescope located at in India an altitude of 4.3 km, and a diameter of 21 m. It has a field of view of around 4$^{\circ}$ and started operation in 2020.\\

{\it MAGIC}\\
MAGIC is made up of two 17 m diameter Cherenkov telescopes, the first of which has operated since 2004 and the second since 2006. They are located in Spain at an altitude of 2.2 km and have a field of view of 3.5$^{\circ}$ \cite{kersz}.\\

{\it SWGO}\\
SWGO is a ground based air shower detector planned for construction in South America. Its location in the Southern hemisphere allows for all sky coverage, and its wide field of view ($\sim$45$^{\circ}$) will make it a very powerful telescope capable of both competing with and supplementing those that already exist. In particular, it aims to improve upon limits above energies of 1 TeV \cite{andrade,hardingswgo}.\\

{\it VERITAS}\\
VERITAS is made up of four 12 m diameter Cherenkov telescopes, operating since 2007 and funded to operate through 2025. It is located in the USA at an altitude of 1.3 km, and has a field of view of 3.5$^{\circ}$ \cite{kersz}. So far it has observed 17 dwarf spheroidals with an exposure time of 630 hours and is able to probe DM in the ultra-heavy range beyond the Planck mass \cite{mcgrath,mcgrath2}.

\subsection{Targets}
\begin{figure}[!h]
    \centering
    \includegraphics[width=0.48\textwidth]{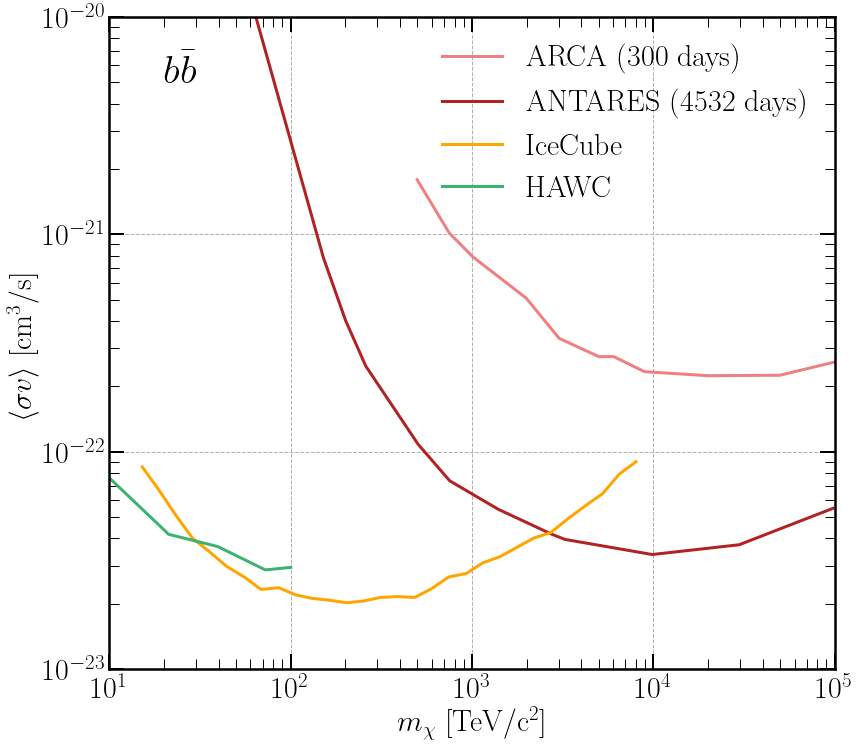}
    \includegraphics[width=0.48\textwidth]{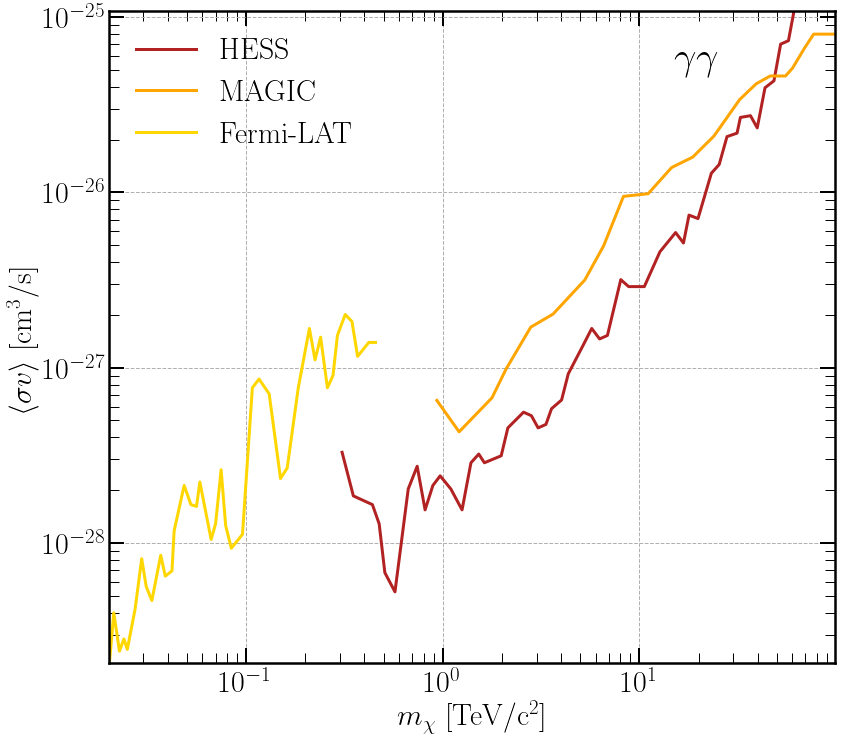}\\
    \includegraphics[width=0.48\textwidth]{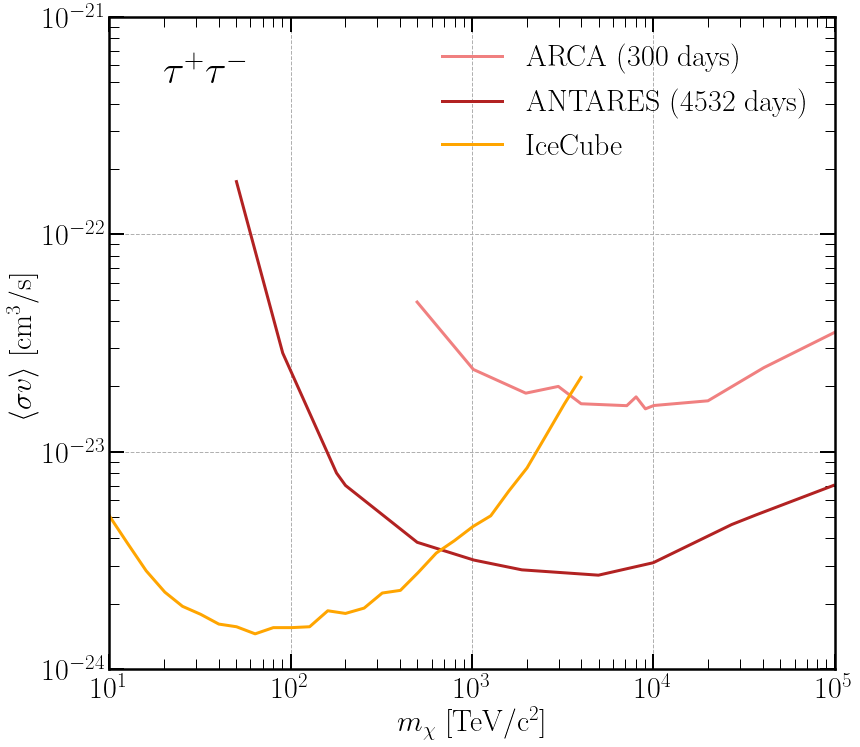}
    \includegraphics[width=0.48\textwidth]{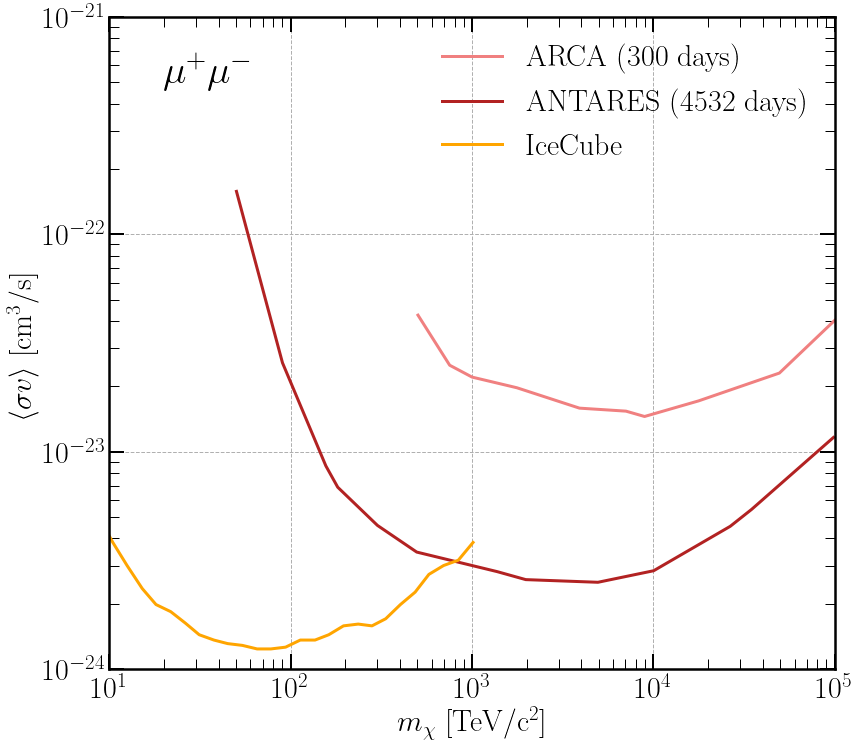}\\
    \includegraphics[width=0.48\textwidth]{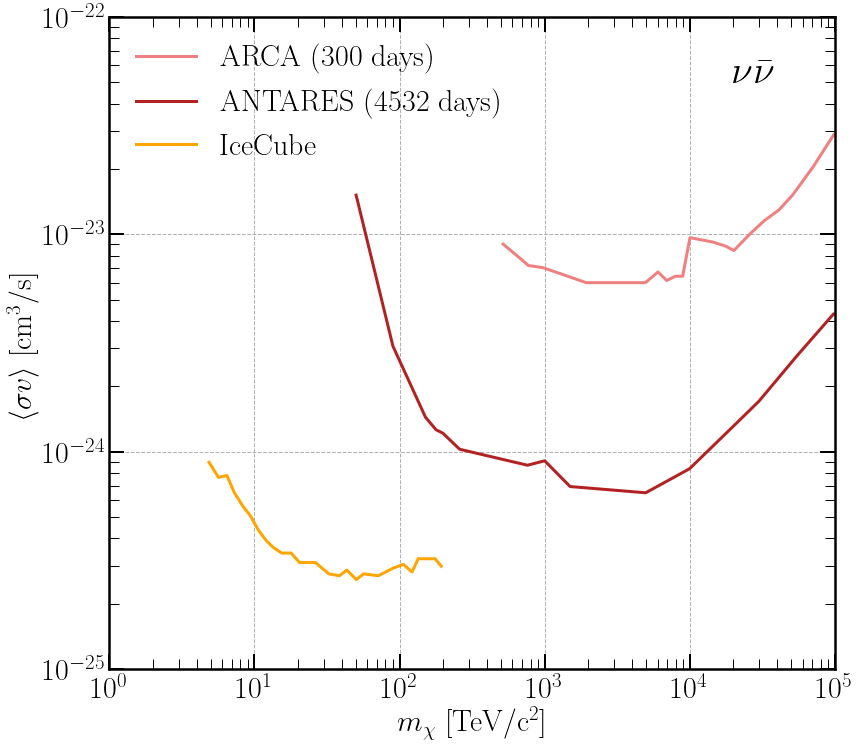}
    \includegraphics[width=0.48\textwidth]{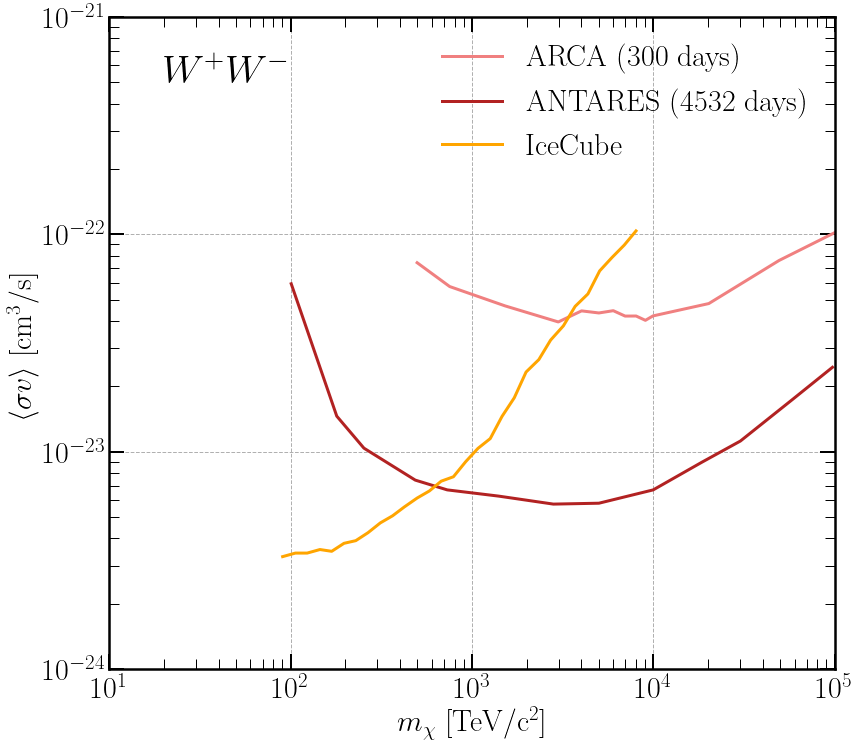}
    \caption{DM limits from the Galactic Centre for different mediator decays (indicated in the top left or right of each plot). ARCA data is from Ref. \cite{saina}, HESS and Fermi-LAT from Ref. \cite{moulin}, MAGIC from Ref. \cite{inada}, IceCube from Ref. \cite{chau}, ANTARES from Ref. \cite{gozzini}, and HAWC from \cite{harding}.}
    \label{fig:centre}
\end{figure}

{\it Galactic Centre}\\
The Galactic Centre makes for a good DM target as it is both nearby and DM dominated, however it is also background contaminated and has an uncertain DM profile. Most Galactic Centre analysis is done using data collected using an ``ON-OFF'' strategy, where data is taken with the telescope pointing towards the Galactic Centre (ON) and away (OFF) \cite{harding,moulin}. The OFF data is assumed to be background only, and can be used to estimate the number of signal events in the ON data set.\\
Typical Galactic Centre searches proceed by assuming DM decays into some mediator, which then decays into Standard Model products that are observed at the detector. Limits from various telescopes are shown in Fig. \ref{fig:centre}. Most final states will produce a broad, decreasing energy spectrum, however where the DM decays or annihilates into two photons directly it will produce a distinctive ``gamma line'' that appears at the mass of the DM. As this is a clear peaked feature, it does not suffer from as much background contamination as other processes \cite{inada,mazziotta}. As well as this, it is possible to set limits to the same scattering cross sections as direct detection by modelling scattering with DM and cosmic rays. These particular limits are competitive in the sub-GeV mass range down to cross sections of 10$^{-31}$ cm$^2$ \cite{reis}.\\

\newpage
{\it Galaxy Clusters}\\
Galaxy clusters provide another astrophysical DM source. They have the benefit of having a high DM content, but are also far away and background contaminated. At ICRC they were used to set limits on both ends of the DM mass spectrum for TeV scale DM and ALPs with masses of 0.1 $\mu$eV. IceCube has a dedicated program to understand if high energy astrophysical neutrinos could be from heavy DM, and places some of the strongest constraints in the TeV-PeV mass range \cite{jeong}. Predicted exclusion limits have also been set in a similar region using CTA projections to examine the excess from the Perseus Cluster \cite{hernandez}. On the other side of the mass scale, MAGIC is also using the Perseus Cluster to place limits on ALPs via their conversion into photons. While there are some hints consistent with the presence of such a conversion, they are not yet statistically significant \cite{damico}.\\

\begin{figure}[!h]
    \centering
    \includegraphics[width=0.49\textwidth]{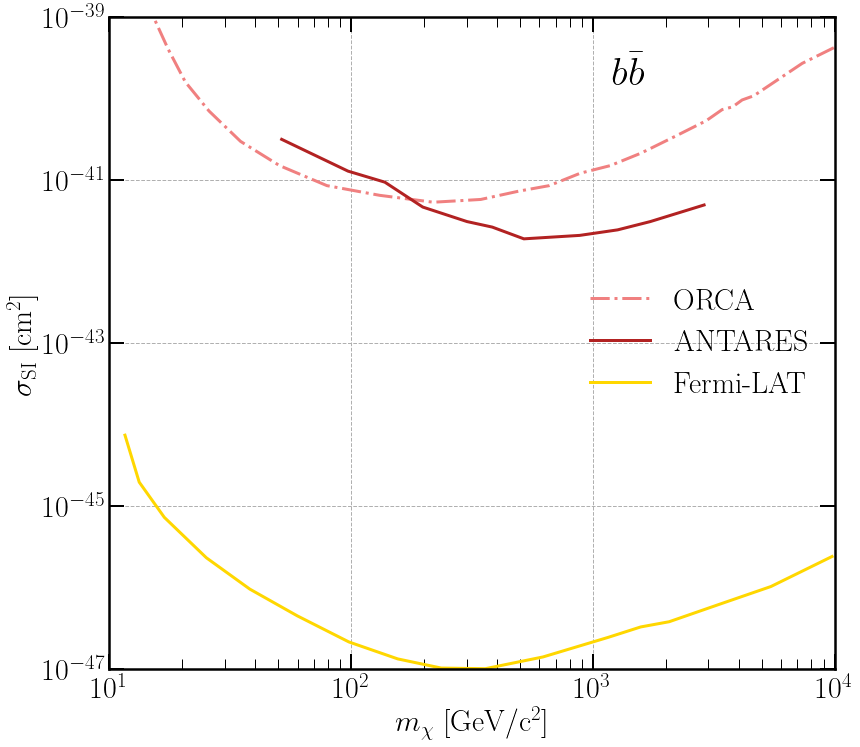}
    \includegraphics[width=0.49\textwidth]{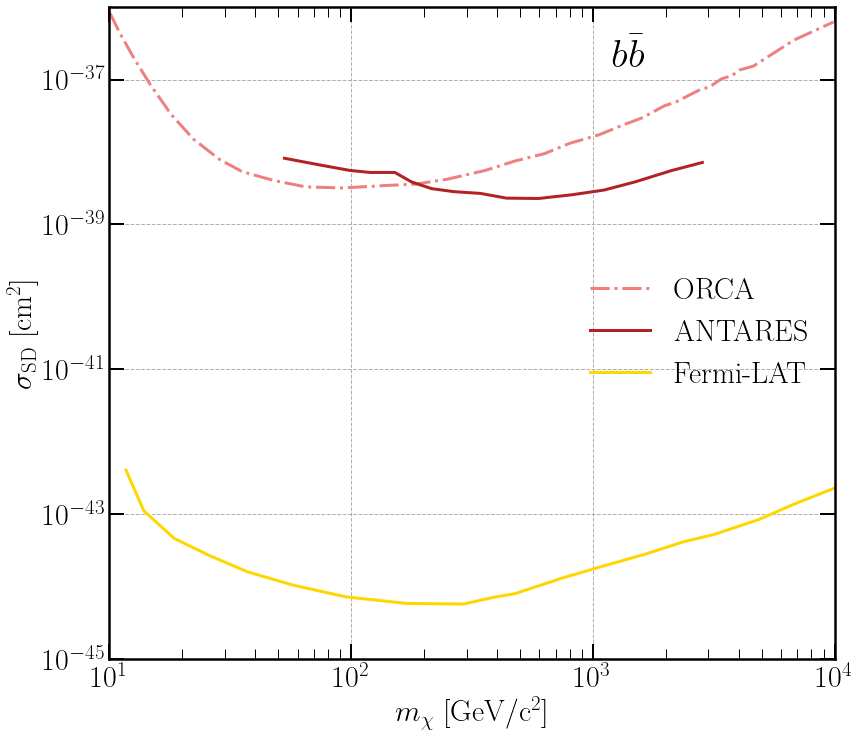}\\
    \includegraphics[width=0.49\textwidth]{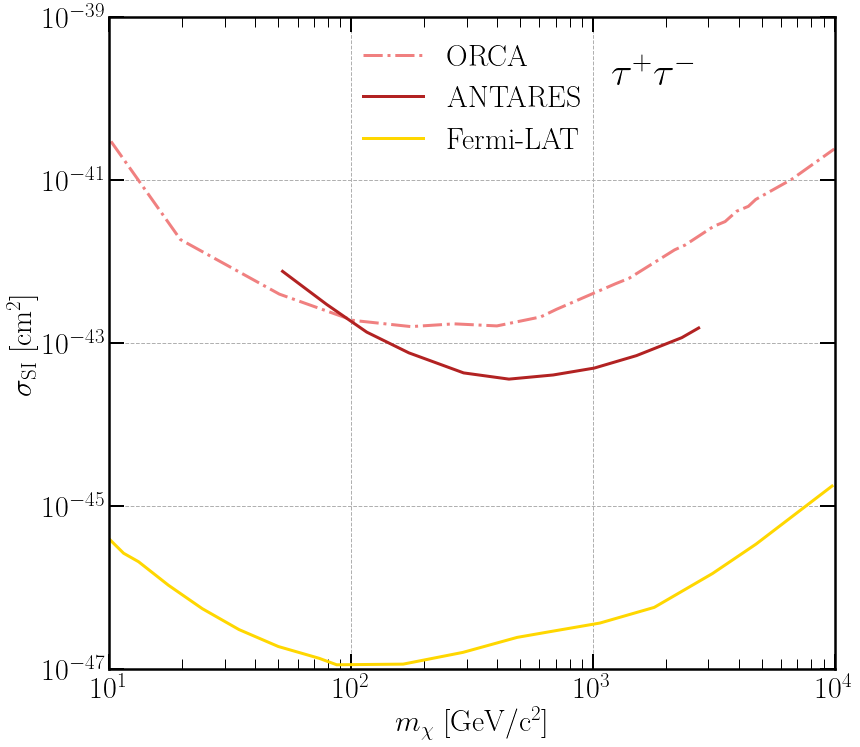}
    \includegraphics[width=0.49\textwidth]{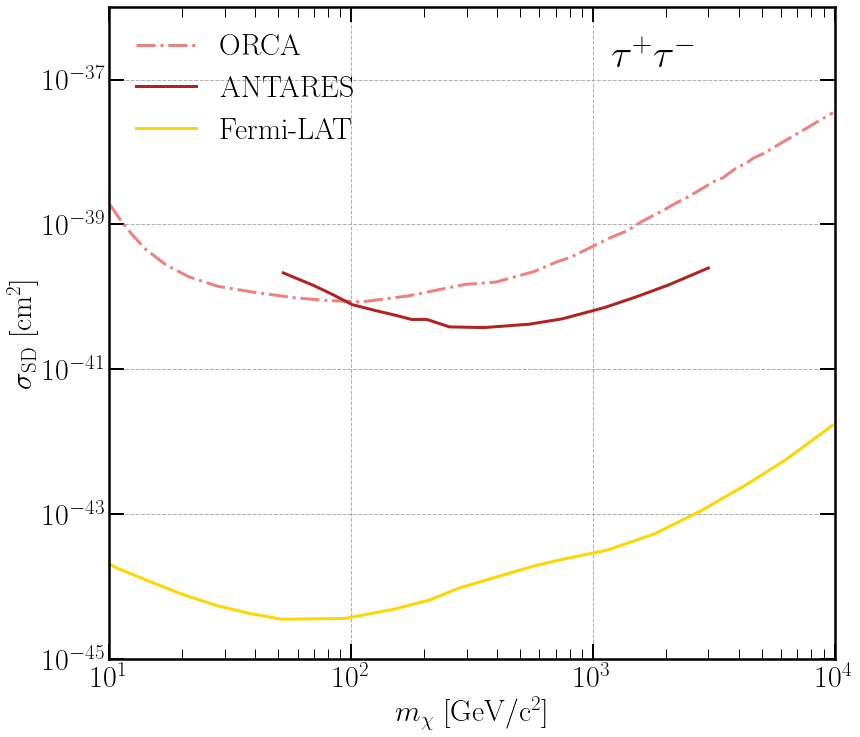}\\
    \includegraphics[width=0.49\textwidth]{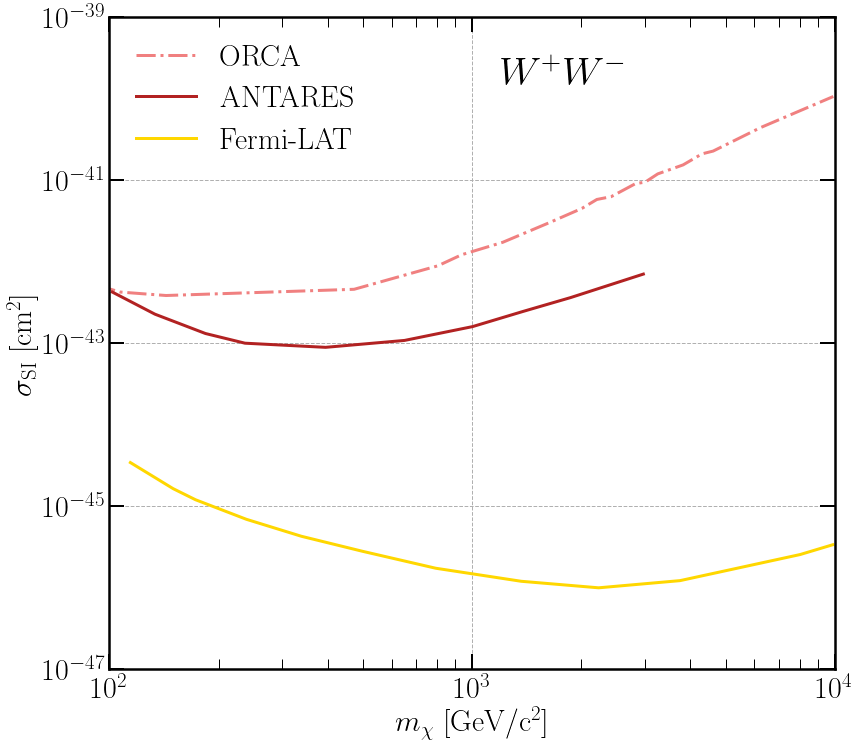}
    \includegraphics[width=0.49\textwidth]{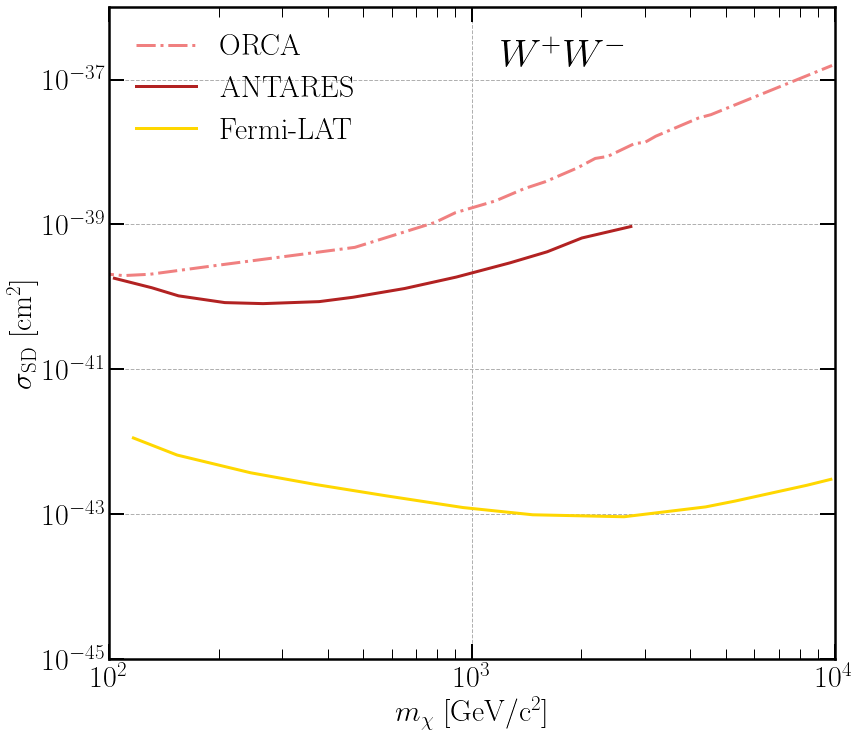}
    \caption{DM limits from solar signals. Limits on a spin independent DM are shown in the left hand column, while those for a spin dependent DM are on the right. The top row are for $b\bar{b}$ decay, the middle $\tau^+ \tau^-$ and bottom $W^+ W^-$. ORCA data is from Ref. \cite{saina, gutierrez}, Fermi-LAT from Ref. \cite{serini}.}
    \label{fig:sol}
\end{figure}

{\it The Sun}\\
DM particles in the galactic halo can be gravitationally trapped in the Sun through interactions with solar nuclei. This can both directly and indirectly produce gamma rays, meaning DM should cause an enhancement of the solar photon flux. The shape of the photon excess will depend on the particular production model; DM annihilation directly into photon pairs will produce line-like features, DM annihilating first into a new light mediator, then into photons will produce a box-shaped excess, and DM annihilating into new light mediators, then into photon via some other mediator (e.g., $b\bar{b}$, $\tau^+\tau^-$) will produce a smooth spectrum. As well as this, by assuming a capture rate of DM in the sun (which will depend on the cross sections typically constrained in direct detection) the solar gamma ray flux can be used to place limits on spin independent or dependent DM comparable to those of direct detection \cite{serini}. These are shown in Fig. \ref{fig:sol} for a subset of possible mediators. Note that the strongest limits (not shown here) are set by Fermi-LAT for the model where the light mediator decays directly to photons $\chi\chi\rightarrow\phi\rightarrow\gamma\gamma$. As well as WIMP DM, it is possible for the Sun to produce solar axions. If they are produced by interactions involving $^{57}$Fe, the resulting signal is a very distinctive energy line, which can be searched for explicitly \cite{onuki}.\\

\newpage

\begin{figure}[!h]
    \centering
    \includegraphics[width=0.49\textwidth]{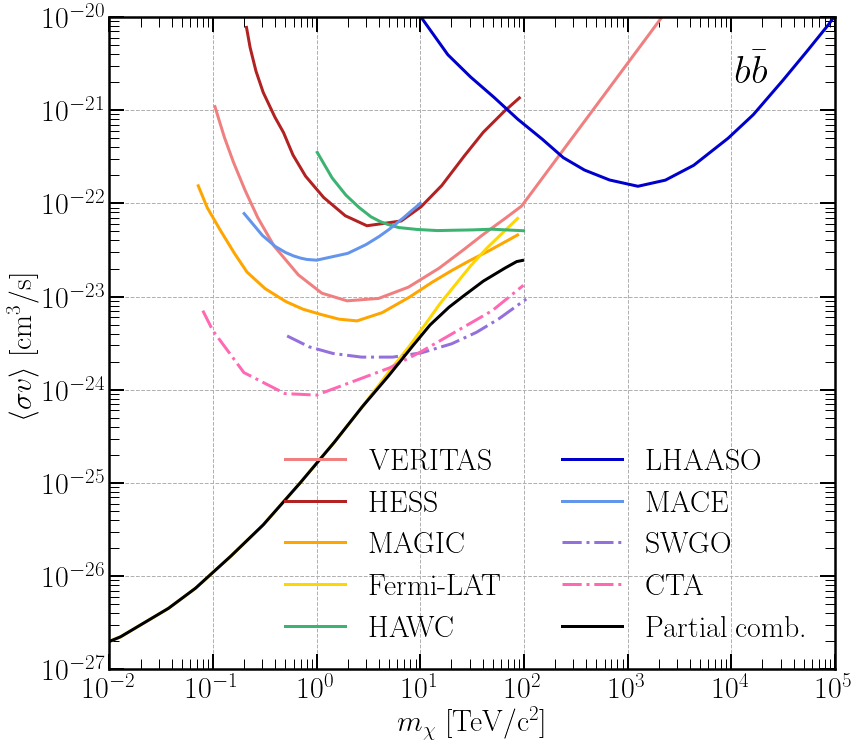}
    \includegraphics[width=0.49\textwidth]{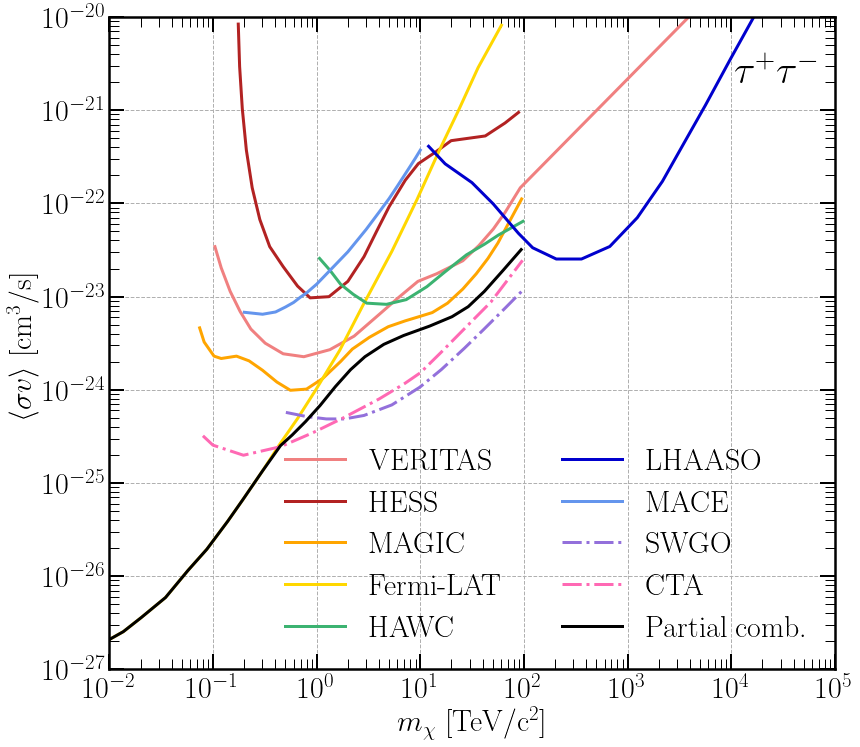}
    \caption{Limits on DM annihilation cross section from dwarf spheroidal searches assuming DM decay into $b\bar{b}$ (left) and $\tau^+ \tau^-$ (right). Dot-dash lines indicated telescope projections, rather than being based on actual data. CTA data is from Ref. \cite{saturni}, VERITAS data from Ref. \cite{mcgrath}, HAWC from \cite{harding}, LHAASO from \cite{li}, SWGO from \cite{andrade}, and MACE data from \cite{khurana}. The ``partial comb.'' refers to the combined analysis presented in Ref. \cite{kersz} that combines data from VERITAS, HESS, MAGIC, Fermi-LAT, and HAWC.}
    \label{fig:dsph}
\end{figure}

\newpage
{\it Dwarf Spheroidals}\\
Dwarf spheroidal galaxies are Milky Way satellites that exhibit a very high mass to light ratio, meaning they have almost no background, but an intrinsically low DM content \cite{saturni}. The key results presented at ICRC 2023 are shown in Fig. \ref{fig:dsph} for the models where DM decays to either $b\bar{b}$ or $\tau^+ \tau^-$. However, it should be noted that the strongest constraints on annihilation cross sections occur for the decay to $\mu^+ \mu^-$, and are set by FAST down to $\langle{\sigma_{{\rm ann}}} v \rangle \sim 10^{-28}$, excluding a DM interpretation of the Galactic Centre Excess for this model \cite{guo}. Another result of note is that the combined analysis of some of the telescopes is able to improve the experimental reach. The method used to compute this can also be extended to other astrophysical messengers in the future \cite{kersz}.

\subsection{Antimatter searches}
As well as typical Standard Model particles, galactic DM decay or annihilation can produce various types of antimatter, in particular composite antimatter such as antihelium, antideuteron and other anti-nuclei, while other new astrophysical sources will not. Thus, an excess of such material would indicate a DM presence in the galaxy. A number of these searches were presented at ICRC.\\
AMS is a spectrometer mounted on the ISS searching for antideuteron in the GeV region. Although a few candidate events have been observed, there are not yet any statistically significant results. Development of the background Monte Carlo is still ongoing, with data to be taken until 2030 \cite{lu}.\\
GAPS is a balloon mission that will be launched from the Antarctic region in 2024 to seach for light cosmic ray anti-nuclei. This is a complicated search requiring careful background modelling, but the experiment is expected to have good sensitivity in the MeV energy region \cite{stoessl}.\\
GRAMS is another balloon experiment probing the MeV energy region. In particular, it is targeting antihelium-3, which provides an almost background free channel to search for \cite{zeng}. However, searches that take advantage of this relatively `clean' signature depend on the survival rate of the antihelium as it travels in the galaxy, which itself depends on inelastic scattering. This can be measured using data from ALICE at the LHC, which has indicated a survival probability of around 50\%, confirming that this is indeed a positive avenue for DM searches \cite{alice}.\\
Finally, CALET is another antimatter detector on board the ISS, though this probes the electron-positron spectrum rather than composite antimatter. Previous limit calculations from both this collaboration and AMS have assumed a single smooth background source. However, in reality this will come from a number of different astrophysical objects. Modelling these more carefully as individual sources gives stricter, more reliable constraints on the DM lifetime \cite{motz}.

\section{Other searches}
There are a number of other astrophysical sources and tests that do not fit neatly into the DM search methods typically discussed in either direct detection or as a traditional, specific source for indirect detection. These will be discussed in this section.\\
Axion conversion into photons is a DM signal getting increasing attention as the WIMP parameter space slowly closes down. This process can produce deviations in the gamma flux observed on Earth, already being probed by HAWC, though significant systematics still need to be understood \cite{pratts, harding}. As well as this, axion-photon conversion can produce high energy gamma rays up to 20 TeV, where typically one would expect attenuation for the gamma ray bursts. There has been a small excess observed that could be explained by this ALP DM, but it is not yet statistically significant enough to make a claim \cite{rojas}.\\
Neutrino point sources will also help to constrain DM. IceCube has developed an agnostic analysis approach that allows for generic interactions of DM and neutrinos to be tested based on attenuation of the observed neutrino flux \cite{wkang}.\\
A similar signal to that caused by axions but due to DM on the opposite mass scale, the decay of various heavy DM can produce a diffuse flux of very high or very low energy gamma rays that could be detected at either LHAASO \cite{nglhaaso} or Fermi-LAT respectively \cite{song}.\\
Although primordial black holes (PBH) appear to be unlikely to explain the full DM population, there is still a small mass window that remains valid. This can be effectively tested using the cross correlation of gamma rays and cosmic microwave background shear. This method produces stricter limits in the mass range of 10$^{14}$ kg (10$^{50}$ eV) \cite{tan}.\\
TeV blazars also give an additional astrophysical insight. They should be producing high energy gammas that remain undetected at telescopes like Fermi-LAT. These ``missing'' events could be explained by ALPs with masses on the scale of 10$^{-10}$-10$^{-9}$ eV \cite{ghosh,jacobsen}.\\
The presence of DM in the galaxy can also produce odd signatures in the transport of particles. For example, DM-proton interactions will impact the transport of cosmic rays in a detectable way, particularly in the GeV-TeV gamma ray spectrum \cite{ambrosone}. Alternatively, if it interacts preferentially with one flavour of neutrinos (e.g., $\nu_{\tau}$), it would impact the flavour ratio on Earth. As this is fairly well constrained, it is possible to place limits on such a DM interaction \cite{katori}. If the DM is slow moving enough, it could also be gravitationally trapped in the Earth, producing a distinctive neutrino signal that would appear to come from the Earth's core, which also has the benefit of probing DM typically too slow to be observed in direct detection experiments \cite{aguilar}.\\
It may also be that indirect detection searches can benefit by casting a wider net in the astrophysical bodies observed. Recent modelling of Omega Centauri has suggested that it may have a dense DM core. Based on this, and its relative proximity to Earth, it would mean that this could provide a better probe of DM than the well studied Dwarf Spheroidals \cite{beck}.\\
DM could also be constrained using synchrotron emission. If DM decays into electrons, it may produce a polarised signal that could be constrained with Planck data. In general, this provides stricter limits than just the intensity of any excess signal alone \cite{manconi}.\\
Finally, there remains the role that colliders can play in DM detection. FASER is a far forward experimental setup to probe neutrinos and other possible long lived particles at the LHC. At present it is able to constrain dark photon mediators \cite{faser}. Upgrades to this experiment are constrained by old infrastructure, and so a new facility, the Forward Physics Facility, has been proposed to support this and other new physics searches \cite{kling}.

\section{Summary}
\begin{figure}[!h]
    \centering
    \includegraphics[width=0.9\textwidth]{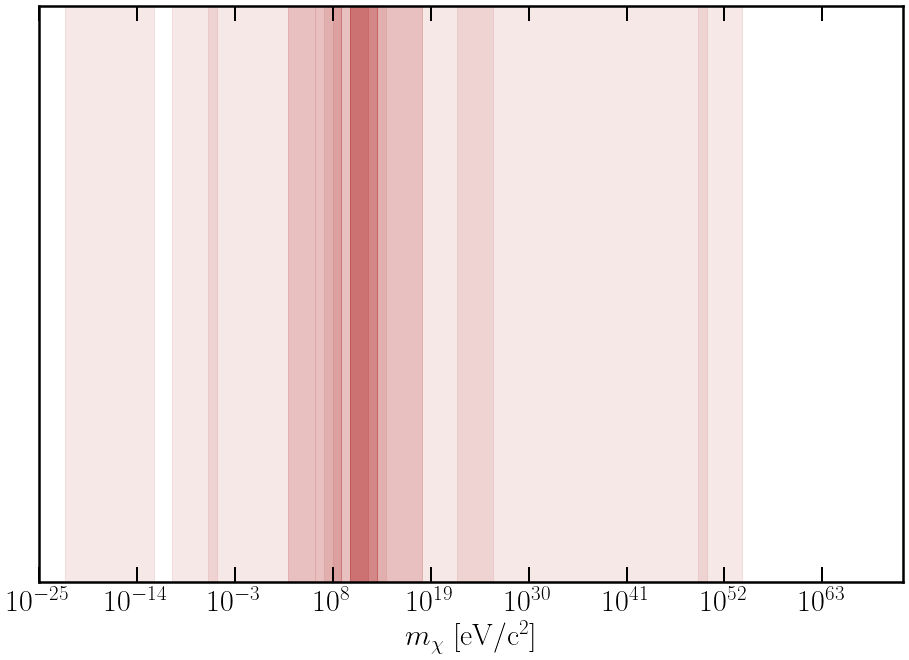}
    \caption{The DM mass phase space constrained at ICRC. Each experimental strategy (e.g., indirect detection with dwarf spheroidals, nuclear recoil direct detection, TeV blazar events) is represented by a red band indicating its sensitivity range. A darker colour means that DM mass can be probed by multiple different strategies. This demonstrates the complementarity of the various methods discussed at the conference.}
    \label{fig:mass_space}
\end{figure}

\begin{figure}[!h]
    \centering
    \includegraphics[width=0.49\textwidth]{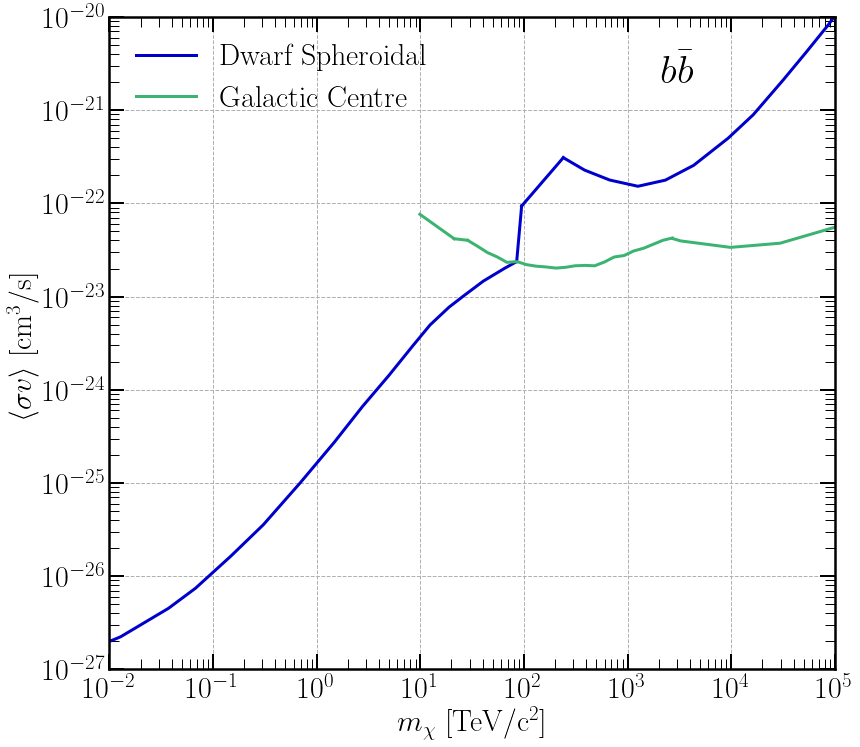}
    \includegraphics[width=0.49\textwidth]{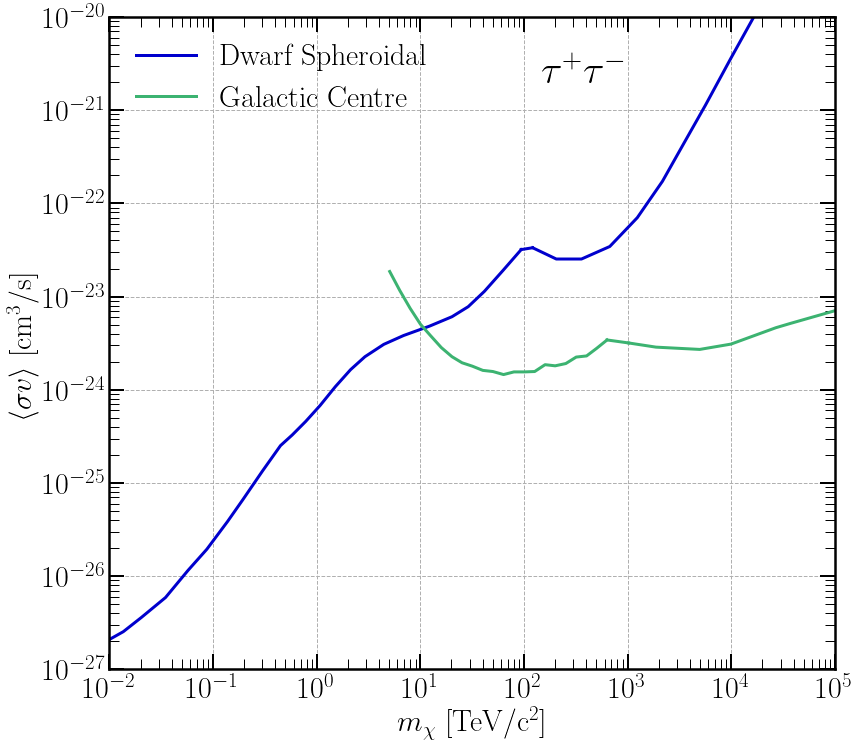}\\
    \includegraphics[width=0.49\textwidth]{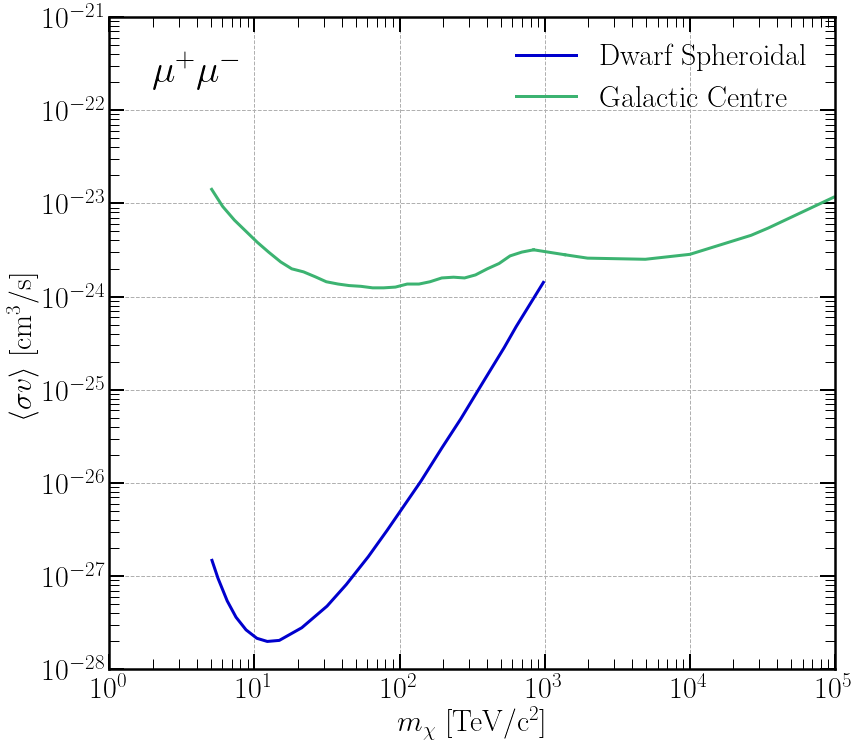}
    \includegraphics[width=0.49\textwidth]{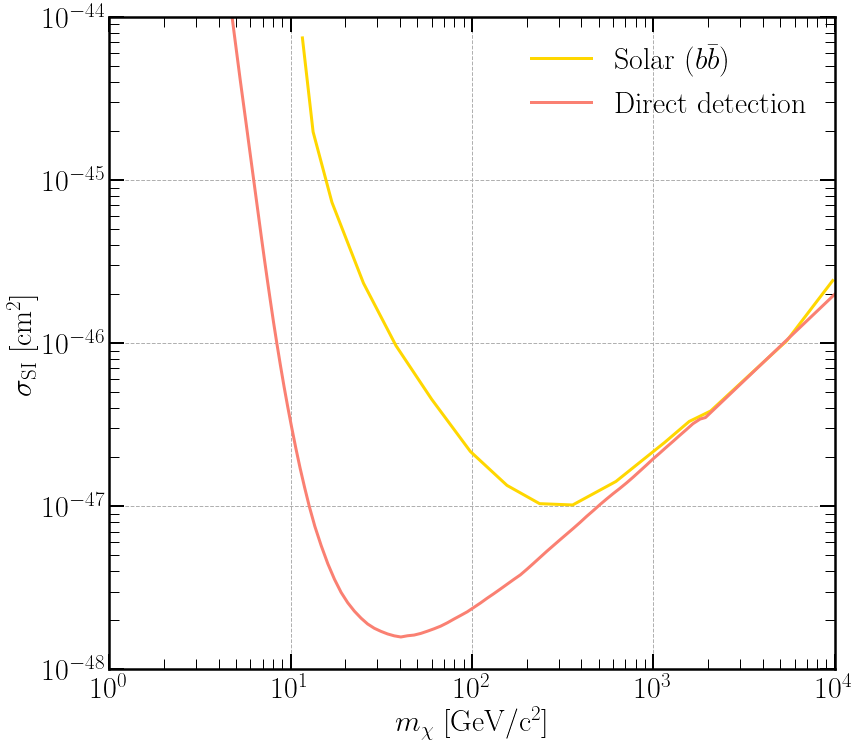}\\
    \caption{Comparison of the best indirect detection limits from dwarf spheroidals and the Galactic Centre with different decay schemes (top row and bottom left). In general the former better constraint the low TeV mass region, while the latter perform better above 10-100 TeV. Note that the dwarf spheroidal limits shown in the bottom left plot come from FAST \cite{guo} and did not appear previously in this paper, as there were no other dwarf spheroidal limits to compare them to under this model assumption.\\
    The bottom right plot demonstrates the comparison of the best direct and indirect limits in typical WIMP region, where (in general) direct detection out performs indirect solar targets for spin independent assumptions. In the spin dependent case, indirect constraints are typically more competitive.}
    \label{fig:comp}
\end{figure}

One of the particularly impressive results from ICRC was the sheer range of mass space that was probed by the experiments discussed, especially those exploring parameter space that they were not explicitly designed to search. Unfortunately, it is difficult to show them all on a single exclusion plot due to the different model assumption requirements for direct and indirect searches. Instead, to demonstrate the depth and complementarity of different methods, the mass range that each search strategy targets is highlighted in red in Fig. \ref{fig:mass_space}. For example, indirect searches targeting dwarf spheroidals is represented by a shaded region from $m_{\chi}=$10$^{10}$-10$^{18}$ eV/c$^2$, while direct detection produces a band from $m_{\chi}=$10$^{8}$-10$^{13}$ eV/c$^2$. Thus, the shade of red demonstrates the number of different methods that are able to probe that mass region - a darker red meaning there are multiple searches targeting that space, while lighter meaning there are only one or two experimental strategies in use. As one might expect, the typical GeV/c$^2$ WIMP mass region is targeted by a large number of experiments, while the higher and lower mass regions are explored with a smaller variety. Note that this {\it only} includes methods and models discussed at ICRC, not the whole field. Where it is possible, a direct comparison of limits from different data sets or search methods is shown in Fig. \ref{fig:comp}.\\
One of the strengths of a conference like ICRC that covers both cosmic rays and DM is the emphasis that one scientist's background is another's signal. Strategies used to clearly identify a cosmic ray signal  are important in background modelling for DM, and vice versa. As detectors scale up in both cost and personnel requirements it is important to squeeze out as much physics as possible, and particularly notable at this conference were experiments reporting limits and results for physics they were not originally designed for, especially neutrino detectors that are being increasingly used to set limits on non-standard DM.\\
Although there was no conclusive DM observations, the presentations at ICRC 2023 have demonstrated impressive strategies, results, and projections, with significant and innovative R\&D for detector design, construction, and analysis promising future upgrades across the DM mass space. A personal highlight for this author was the emphasis on global collaboration, open data, and combined analysis. No matter how powerful the detector, there is no way for a single experiment to understand DM if (or hopefully when) we see a signal. As such, it is good to see the community moving towards a more inter-collaborative approach, utilising the impressive complementarity of the various detection methods discussed.

\newpage
\section*{Acknowledgements}
I would like to thank the organisers of the conference for inviting me as the dark matter rapporteur, and the participants for a fascinating series of talks and scintillating discussions.


\newpage
\begin{thebibliography}{99}

\bibitem{calore}
F. Calore, Dark matter searches: status and prospects, {\it PoS}{\bf (ICRC2023)}025

\bibitem{skang}
S. Kang, Halo-independent bounds on the WIMP-nucleon couplings from direct detection and neutrino observations based on the non-relativistic effective theory, {\it PoS}{\bf (ICRC2023)}1371

\bibitem{mcgrath}
C. McGrath, An indirect search for Dark Matter with a combined analysis of dwarf spheroidal galaxies from VERITAS, {\it PoS}{\bf (ICRC2023)}1395

\bibitem{kersz}
D. Kerszberg, Search for dark matter annihilation with a combined analysis of dwarf spheroidal galaxies from Fermi-LAT, HAWC, H.E.S.S., MAGIC and VERITAS, {\it PoS}{\bf (ICRC2023)}1426

\bibitem{eckner}
C. Eckner, The impact of model realism on interpretations of the Galactic Center Excess, {\it PoS}{\bf (ICRC2023)}1401

\bibitem{batista}
R. Alves Batista et. al, Axion-like particles and high-energy gamma rays: interconversion revisited, {\it PoS}{\bf (ICRC2023)}1383

\bibitem{aguirre}
A. Aguirre-Santaella, Shedding light on low-mass subhalo survival and annihilation luminosity with numerical simulations, {\it PoS}{\bf (ICRC2023)}004

\bibitem{ngnu}
K. C. Y. Ng, Towards detecting super-GeV dark matter via annihilation to neutrinos, {\it PoS}{\bf (ICRC2023)}1451

\bibitem{desimone}
M. Giovanna Dianotti et. al, Supernovae Ia and Gamma-Ray Bursts together shed new lights on the Hubble constant tension and cosmology, {\it PoS}{\bf (ICRC2023)}1367

\bibitem{avila}
I. Maturana Avila, An extensive study of the scotogenic model with scalar dark matter, {\it PoS}{\bf (ICRC2023)}1368


\bibitem{yu}
G. Yu, Dark matter search using NaI(Tl) at the COSINE-100 experiment, {\it PoS}{\bf (ICRC2023)}1421

\bibitem{buttazzo}
D. Buttazzo et. al, Annual modulations from secular variations: relaxing DAMA?, {\bf{JHEP2020}}1347

\bibitem{messina}
A. Messina, M. Nardecchia, S. Piacentini, Annual modulations from secular variations: not relaxing DAMA?, {\bf{JCAP2020}}{\it (04)}037

\bibitem{mews}
M. Mews, The SABRE South Experiment at the Stawell Underground Physics Laboratory, {\it PoS}{\bf (ICRC2023)}1370

\bibitem{fushimi}
K. Fushimi et. al, PICOLON dark matter search project, {\it PoS}{\bf (ICRC2023)}1404

\bibitem{kotera}
K. Kotera, Result and analysis for Ingot\#94 of PICOLON ultra-pure NaI(Tl) crystal, {\it PoS}{\bf (ICRC2023)}1439

\bibitem{tardif}
R. Turcotte-Tardif, DEAP-3600 - Latest results from the largest liquid argon dark matter experiment, {\it PoS}{\bf (ICRC2023)}1388

\bibitem{walczak}
M. Walczak, Searching for dark matter with liquid-argon detectors, {\it PoS}{\bf (ICRC2023)}1425

\bibitem{wang}
J. R. Wang, Status and results of the LZ experiment, {\it PoS}{\bf (ICRC2023)}1409

\bibitem{eising}
H. Schulze Eising, Radon removal in the XENONnT experiment via cryogenic distillation, {\it PoS}{\bf (ICRC2023)}1405

\bibitem{kobayashi}
M. Kobayashi, The purification system for the XENONnT dark matter search experiment, {\it PoS}{\bf (ICRC2023)}1423

\bibitem{brown}
A. Brown, XENONnT: status of the experiment and latest results, {\it PoS}{\bf (ICRC2023)}1379

\bibitem{gaior}
R. Gaior et. al, XeLab: a test platform for xenon TPC instrumentation, {\it PoS}{\bf (ICRC2023)}1420

\bibitem{hasegawa}
T. Hasegawa, Development of a hybrid-photodetector for the DARWIN experiment, {\it PoS}{\bf (ICRC2023)}1432

\bibitem{sakamoto}
S. Sakamoto et. al, Development of a low-noise SiPM for the DARWIN experiment, {\it PoS}{\bf (ICRC2023)}1435

\bibitem{ito}
H. Ito, Screening ultra-low alpha emissivity from the material surface based on a gaseous TPC with PMTs, {\it PoS}{\bf (ICRC2023)}1374

\bibitem{botti}
A. M. Botti, The SENSEI Experiment: sub-GeV dark matter searches with skipper-CCDs, {\it PoS}{\bf (ICRC2023)}1398

\bibitem{erhart}
A. Erhart, At the 100eV Frontier: Calibrating Nuclear Recoils with CRAB, {\it PoS}{\bf (ICRC2023)}1418

\bibitem{damic}
R. Gaior, The DAMIC-M experiment: scientific results from prototype detector and development status, {\it PoS}{\bf (ICRC2023)}1419

\bibitem{oscura}
A. M. Botti and C. Chavez, Novel multi-channel skipper-CCD packages for dark matter searches, {\it PoS}{\bf (ICRC2023)}1397

\bibitem{higashino}
S. Higashino, NEWAGE: direction-sensitive direct dark matter search, {\it PoS}{\bf (ICRC2023)}1434

\bibitem{paun}
A. Păun, G.E. Păvălaş and V. Popa, KM3NeT sensitivity to a flux of down-going nuclearites, {\it PoS}{\bf (ICRC2023)}1382

\bibitem{shinozaki}
K. Shinozaki et. al, Status of the DIMS project for macroscopic dark matter search using ultra-high sensitivity CMOS cameras at the Telescope Array UHECR observatory, {\it PoS}{\bf (ICRC2023)}1390

\bibitem{mori}
M. Mori, Power supply system construction and power monitoring analysis for the dark matter search experiment DIMS, {\it PoS}{\bf (ICRC2023)}1392

\bibitem{kajino}
F. Kajino, DIMS Experiment for Macroscopic Dark Matter and Interstellar Meteoroid Study, {\it PoS}{\bf (ICRC2023)}1376

\bibitem{casolino}
M. Casolino, Search for Strange Quark Matter from the International Space Station with the Mini-EUSO experiment, {\it PoS}{\bf (ICRC2023)}1410

\bibitem{iovine}
N. Iovine, Search for Boosted Dark Matter in Super-Kamiokande with low energy electrons, {\it PoS}{\bf (ICRC2023)}1365

\bibitem{nozzoli}
F. Nozzoli, Search for light fermionic Dark Matter with Double Beta decay experiments., {\it PoS}{\bf (ICRC2023)}1417

\bibitem{garcmend}
J. Garcia-Mendez, S. Ardid and M. Ardid, Dark matter search towards the Sun using Machine Learning reconstructions of single-line events in ANTARES, {\it PoS}{\bf (ICRC2023)}1443

\bibitem{gozzini}
S. R. Gozzini and J. D. D. Zornoza, Dark matter searches with the full data sample of the ANTARES neutrino telescope, {\it PoS}{\bf (ICRC2023)}1375

\bibitem{harding}
P. Harding on behalf of the HAWC Collaboration, Beyond the Standard Model with HAWC, {\it PoS}{\bf (ICRC2023)}1400

\bibitem{saina}
A. Saina, Indirect Search for Dark Matter with the KM3NeT Neutrino Telescope, {\it PoS}{\bf (ICRC2023)}1377

\bibitem{gutierrez}
M. Gutierrez, A. Saina, S. Navas and S.R. Gozzini, Search for dark matter towards the Sun with the KM3NeT/ORCA6 neutrino telescope, {\it PoS}{\bf (ICRC2023)}1406

\bibitem{li}
J. Li, Dark Matter Limits from Dwarf Spheroidal Galaxies with the LHAASO-KM2A, {\it PoS}{\bf (ICRC2023)}1381

\bibitem{andrade}
M. Andrade on behalf of The SWGO Collaboration, Dark Matter searches in Dwarf Galaxies with the Southern Wide-field Gamma-ray Observatory, {\it PoS}{\bf (ICRC2023)}1413

\bibitem{hardingswgo}
P. Harding on behalf of The SWGO Collaboration, Beyond the Standard Model with the Southern Wide-field Gamma-ray Observatory, {\it PoS}{\bf (ICRC2023)}1399

\bibitem{mcgrath2}
C. McGrath, Indirect dark matter search beyond the unitarity limit with VERITAS, {\it PoS}{\bf (ICRC2023)}1450

\bibitem{moulin}
A. Montanari and E. Moulin, Search for dark matter gamma-ray line annihilation signals in the H.E.S.S. Inner Galaxy Survey, {\it PoS}{\bf (ICRC2023)}1424

\bibitem{inada}
T. Inada on behalf of the MAGIC Collaboration, Gamma-ray Spectral Line emission search from Dark Matter Annihilation up to 100 TeV towards the Galactic Centre with MAGIC, {\it PoS}{\bf (ICRC2023)}1427

\bibitem{mazziotta}
 M.N. Mazziotta, M. Giliberti, F. Loparco and D. Serini, Search for Dark Matter Lines in the gamma-ray energy spectra with the Fermi-LAT, {\it PoS}{\bf (ICRC2023)}1387

 \bibitem{reis}
 I. Reis, E. Moulin and A. Viana, Indirect detection of inelastically scattered Dark Matter, {\it PoS}{\bf (ICRC2023)}1408

 \bibitem{chau}
N. T Chau and J. Augilar on behalf of the IceCube Collaboration, Indirect dark matter search in the Galactic Centre with IceCube, {\it PoS}{\bf (ICRC2023)}1394

\bibitem{jeong}
M. Jeong and C. Rott on behalf of the IceCube Collaboration, Search for Dark Matter Decay in Nearby Galaxy Clusters and Galaxies with IceCube, {\it PoS}{\bf (ICRC2023)}1378

\bibitem{hernandez}
S. Hernandez-Cadena on behalf of the CTA Collaboration, Expected exclusion limits to TeV dark matter from the Perseus Cluster with the Cherenkov Telescope Array, {\it PoS}{\bf (ICRC2023)}1436

\bibitem{damico}
G. D'Amico on behalf of the MAGIC Collaboration, Constraints to axion-like particles with the Perseus Galaxy Cluster with MAGIC, {\it PoS}{\bf (ICRC2023)}1442

\bibitem{serini}
D. Serini, M. Giliberti, F. Loparco and M. Nicola Mazziotta, Dark matter searches with solar gamma rays using the Fermi-LAT, {\it PoS}{\bf (ICRC2023)}1384

\bibitem{onuki}
Y. Onuki, ``ISAI'' Investigating Solar Axion by Iron-57, {\it PoS}{\bf (ICRC2023)}1389

\bibitem{saturni}
F. G. Saturni on behalf of The CTA Consotrium, Dark matter searches in dwarf spheroidal galaxies with the Cherenkov Telescope Array, {\it PoS}{\bf (ICRC2023)}1366

\bibitem{guo}
W. Q. Guo, Constraints on dark matter annihilation from the FAST observation of the Coma Berenices dwarf galaxy, {\it PoS}{\bf (ICRC2023)}1411

\bibitem{khurana}
M. Khurana, Prospects of detecting gamma-ray signal of dark matter interaction with the MACE telescope, {\it PoS}{\bf (ICRC2023)}1412

\bibitem{lu}
S. Lu, Cosmic Anti-Deuterons measured with the Alpha Magnetic Spectrometer on the ISS, {\it PoS}{\bf (ICRC2023)}1391

\bibitem{stoessl}
A. Stoessel, The GAPS experiment - a search for light cosmic ray antinuclei, {\it PoS}{\bf (ICRC2023)}1440

\bibitem{zeng}
J. Zeng, AntiHelium-3 Search with the GRAMS Experiment, {\it PoS}{\bf (ICRC2023)}1407

\bibitem{alice}
L. Šerkšnytė, First measurement of the antihelium-3 inelastic cross section and its implications for indirect dark matter searches, {\it PoS}{\bf (ICRC2023)}1372

\bibitem{motz}
H. Motz on behalf of the CALET collaboration, Dark Matter Limits from the CALET Electron$+$Positron Spectrum with Individual Astrophysical Source Background, {\it PoS}{\bf (ICRC2023)}1385

\bibitem{pratts}
A. Pratts on behalf of the HAWC Collaboration, ALPs searches with galactic sources using the HAWC Observatory, {\it PoS}{\bf (ICRC2023)}1417

\bibitem{rojas}
D. Avila, Constrains on the ALPs parameter space from GRB 221009A TeV emission, {\it PoS}{\bf (ICRC2023)}1438

\bibitem{wkang}
W. Kang on behalf of the IceCube Collaboration, Search for the rare interactions of neutrinos from distant point sources with the IceCube Neutrino Telescope, {\it PoS}{\bf (ICRC2023)}1380

\bibitem{nglhaaso}
K. C. Y. Ng, Constraints on heavy decaying dark matter from 570 days of LHAASO observations, {\it PoS}{\bf (ICRC2023)}1444

\bibitem{song}
D. Song, Searching for cascaded gamma rays from very heavy dark matter with the Fermi LAT, {\it PoS}{\bf (ICRC2023)}1416

\bibitem{tan}
X. H. Tan et. al, Searching for Primordial Black Hole as Dark matter by cross-correlation with MeV $\gamma$-ray emissions, {\it PoS}{\bf (ICRC2023)}1364

\bibitem{ghosh}
O. Ghosh and S. Bhattacharyya, Light Dark Matter in a Blazar-heated Universe, {\it PoS}{\bf (ICRC2023)}1448

\bibitem{jacobsen}
S. Jacobsen et. al, Axion-like particles as a solution to an emerging tension in the TeV gamma-ray sky, {\it PoS}{\bf (ICRC2023)}1446

\bibitem{ambrosone}
A. Ambrosone et. al, Constraining Sub-GeV Dark Matter through Proton-Dark Matter Scatterings in Starburst Nuclei, {\it PoS}{\bf (ICRC2023)}1386

\bibitem{katori}
T. Katori, C. Argüelles and K.R. Farrag, Ultra-light Dark Matter Limits from Astrophysical Neutrino Flavour, {\it PoS}{\bf (ICRC2023)}1415

\bibitem{aguilar}
J. A. Aguilar Sanchez on behalf of the IceCube Collaboration, Search for Dark Matter annihilation in the center of the Earth with IceCube, {\it PoS}{\bf (ICRC2023)}1393

\bibitem{beck}
G. Beck, Multi-frequency dark matter searches in Omega Centauri, {\it PoS}{\bf (ICRC2023)}1414

\bibitem{manconi}
S. Manconi, Galactic dark matter constrained by synchrotron emission, {\it PoS}{\bf (ICRC2023)}1441

\bibitem{faser}
T. Inada, Looking Forward to New Physics at Dark Sectors with the FASER experiment at the LHC, {\it PoS}{\bf (ICRC2023)}1428

\bibitem{kling}
F. Kling, The Forward Physics Facility and its Implications for Astroparticle Physics, {\it PoS}{\bf (ICRC2023)}023



\end{thebibliography}
\end{document}